\documentclass[10pt,journal,compsoc]{IEEEtran}
\PassOptionsToPackage{table}{xcolor}
\usepackage{multirow}
\usepackage[skins]{tcolorbox}

\usepackage{subcaption}
\usepackage{xspace}

\usepackage{amsmath,amssymb}
\usepackage{booktabs}
\usepackage{xspace}
\usepackage{multirow}
\usepackage{balance}
\usepackage{paralist}
\usepackage{microtype}
\usepackage{url}
\usepackage{listings}
\usepackage{graphicx}

\usepackage{tikz}
\usepackage{xcolor}

\usepackage{float}
\newtcolorbox{custombox}[1]{
	colback=gray!10,
	colframe=gray!20,
	left=1mm,
	right=1mm,
	top=1mm,
	bottom=1mm,
	fonttitle=\bfseries,
	arc=2mm,
	leftrule=0mm,
	rightrule=.5mm,
	toprule=0mm,
	bottomrule=.5mm,
	notitle,
	before=\par\smallskip\noindent,
	before upper={\textbf{#1: } },
}

\newsavebox\CBox

\ifCLASSOPTIONcompsoc
  \usepackage[nocompress]{cite}
\else
  \usepackage{cite}
\fi
\ifCLASSINFOpdf
\else
\fi

\usepackage{flushend}
\usepackage{booktabs}
\usepackage{xspace}
\usepackage{multirow}
\usepackage{balance}
\usepackage{paralist}
\usepackage{microtype}
\usepackage{listings}
\usepackage{graphicx}

\lstdefinestyle{javastyle}{
    backgroundcolor=\color{white},   
    commentstyle=\color{teal},
    keywordstyle=\color{purple},
    numberstyle=\tiny\color{gray},
    basicstyle=\ttfamily\scriptsize,
    breakatwhitespace=false,         
    breaklines=true,                 
    captionpos=b,                    
    keepspaces=true,                 
    numbers=left,                    
    numbersep=5pt,                  
    showspaces=false,                
    showstringspaces=false,
    showtabs=false,                  
    tabsize=1
}
\usepackage{float}
\usepackage[ruled, vlined, linesnumbered]{algorithm2e}
\SetAlFnt{\small}
\SetAlCapNameFnt{\small}
\SetAlCapFnt{\small}
\SetFuncSty{textsc}
\AtBeginEnvironment{algorithm}{
	\DontPrintSemicolon
}

\newtheorem{defin}{Definition}

\newcommand{\chunk}[2]{%
	\fcolorbox{black}{yellow}{\bfseries\sffamily\scriptsize#1}%
	{$\blacktriangleright$#2$\blacktriangleleft$}%
}

\RequirePackage{float}
\RequirePackage[skins]{tcolorbox}

\newtcolorbox{probdefinition}[1]{
	colback=gray!10,
	colframe=gray!22,
	left=1mm,
	right=1mm,
	top=1mm,
	bottom=1mm,
	fonttitle=\bfseries,
	arc=2mm,
	leftrule=0mm,
	rightrule=1mm,
	toprule=0mm,
	bottomrule=1mm,
	notitle,
	before=\par\smallskip\noindent,
	before upper={\textbf{#1: } },
}

\newcommand{\paolo}[1]{\chunk{Paolo}{\textbf{\textcolor{red}{\textsl{#1}}}}}

\newcommand{\aitor}[1]{\chunk{Aitor}{\textbf{\textcolor{red}{\textsl{#1}}}}}
\newcommand{\jon}[1]{\chunk{Jon}{\textbf{\textcolor{red}{\textsl{#1}}}}}

\newcommand{\tool}{\textsc{Gen\-Morph}\@\xspace}
\newcommand{\gassert}{\textsc{GAs\-sert}\@\xspace}

\AtBeginDocument{%
  \providecommand\BibTeX{{%
    \normalfont B\kern-0.5em{\scshape i\kern-0.25em b}\kern-0.8em\TeX}}}

\begin{document}
%
\title{\tool: Automatically Generating Metamorphic Relations via Genetic Programming}%
%
%
%

\author{Jon~Ayerdi, 
        Valerio~Terragni, 
        Gunel~Jahangirova, 
        Aitor~Arrieta, 
        and~Paolo~Tonella
\IEEEcompsocitemizethanks{\IEEEcompsocthanksitem J. Ayerdi and A. Arrieta are with the Department
of Electrical and Computer Engineering, Mondragon University, Spain.\protect\\
E-mail: jayerdi@mondragon.edu
\IEEEcompsocthanksitem V. Terragine is with Auckland University, New Zealand
\IEEEcompsocthanksitem G. Jahangirova is with the King's College London, United Kingdom
\IEEEcompsocthanksitem P. Tonella is with the Università della Svizzera italiana (USI), Switzerland
}
}

\IEEEtitleabstractindextext{%
\begin{abstract}
Metamorphic testing is a popular approach that aims to alleviate the oracle problem in software testing. At the core of this approach are Metamorphic Relations (MRs), specifying properties that hold among multiple test inputs and corresponding outputs.
Deriving MRs is mostly a manual activity, since their automated generation is a challenging and largely unexplored problem.

This paper presents \tool, a technique to automatically generate MRs for Java methods that involve inputs and outputs that are boolean, numerical, or ordered sequences. 
\tool uses an evolutionary algorithm to search for \textit{effective} test oracles, i.e., oracles that trigger no false alarms and expose software faults in the method under test. The proposed search algorithm is guided by two fitness functions that measure the number of false alarms and the number of missed faults for the generated MRs.
Our results show that \tool generates effective MRs for 18 out of 23 methods (mutation score \textgreater20\%). Furthermore, it can increase \textsc{Randoop}'s fault detection capability in 7 out of 23 methods, and \textsc{Evosuite}'s in 14 out of 23 methods. When compared with \textsc{AutoMR}, a state-of-the-art MR generator, \tool also outperformed its fault detection capability in 9 out of 10 methods.

\end{abstract}

}

\maketitle

\IEEEdisplaynontitleabstractindextext

%
\IEEEpeerreviewmaketitle


\section{Introduction}
The increasing complexity of software systems makes software test automation both increasingly important and challenging. 
One of the main issues in this domain is the automation of the test oracle problem, the problem of distinguishing between correct and incorrect test executions~\cite{barr2014oracle}.
While recent years have witnessed significant advances in automated test input generation~\cite{fraser2011evosuite,pacheco2007randoop,lukasczyk2020automated,mcminn2004search,panichella2017automated}, the oracle problem remains one of the main bottlenecks to achieve full test automation~\cite{harman2013comprehensive}.

\textbf{Metamorphic Testing~(MT)}~\cite{1998-chen-tr} aims at alleviating the oracle problem. 
It is based on the intuition that even if we cannot automatically determine the correctness of the output for an individual input, it may still be possible to use relations among the expected outputs of multiple related inputs as oracles~\cite{2017-chen-cs}.
Existing research on metamorphic testing has proven that such relations, called \textbf{Metamorphic Relations~(MRs)}, exist in virtually any non-trivial software system~\cite{2016-segura-tse}.

For example, consider a program \texttt{f} that computes the square of a number. We cannot determine the correctness of \texttt{f(5)} without knowing that the expected value is \texttt{25}. However, a mathematical property of the square function is $x^2 = (-x)^2$. This property can be represented as an MR that verifies that the square functions of two numbers (one the negation of the other) are identical.
If two inputs violate this MR, we have found a fault in the implementation of \texttt{f}.

The key advantage of MT is that a single MR can be applied to multiple test inputs. This is generally not true for canonical test assertions that often predicate on specific test inputs. For example, the test assertion \texttt{assertEquals(16,f(4))} is specific to the input \texttt{4}.
Manually deriving assertions for thousands of automatically generated test inputs is infeasible.
Indeed, test generators often need many test inputs to be effective~\cite{comparison}. Conversely, a single MR can be applied to different test inputs, as long as they satisfy the input relation.

With the rapid advance of test input generators, the popularity of metamorphic testing has drastically increased~\cite{2016-segura-tse,2017-chen-cs,chen2021new}.
Recently, large companies such as  \textsc{Meta}~\cite{ahlgren2021testing}, \textsc{Google}~\cite{graphicsfuzz}, \textsc{Adobe}~\cite{2017-jarman-met}, and NASA~\cite{2015-lindvall-icse} are leveraging MT for testing their software systems. 
While the usefulness of MT is well recognized among both researchers and
the industry~\cite{chen2021new}, identifying MRs largely remains a costly manual activity that requires domain knowledge~\cite{2010-chen-sose}. 
The automated discovery of MRs is an important research topic in MT, as it would reduce the human cost associated with metamorphic testing and enable full test automation~\cite{2016-segura-tse,2017-chen-cs,chen2021new}.
Unfortunately, it is a challenging problem that remains largely unexplored~\cite{automr,2016-segura-tse,2017-chen-cs,chen2021new,2014-zhang-ase}. 

\smallskip
This paper presents \textbf{\tool} (\emph{\underline{\textbf{Gen}}erator of Meta\underline{\textbf{morph}}ic Relations}) a technique to automatically discover MRs for Java methods. 
\tool relies on search-based software engineering (in particular, genetic programming) to explore the space of candidate solutions, driven by a fitness function that rewards MRs with fewer false positives and false negatives. 
In the context of test oracles~\cite{oasis}, \textit{false positives} represent correct program executions in which the oracle fails but should pass, and \textit{false negatives} represent incorrect program executions in which the oracle passes but should fail~\cite{oasis,oasistse,oasistool}.
Correspondingly, a high-quality MR is one with no false alarms (zero false positives) that is useful to expose software faults (few false negatives).

We evaluated \tool on 23 Java methods from three different open-source libraries.
The results show that \tool generates effective MRs for 18 of these subjects. Furthermore, the generated MRs increase the fault detection capability of automatically generated regression test suites for 14 subjects. Compared with \textsc{AutoMR}, a state-of-the-art MR generation tool, \tool achieved a higher fault detection capability with 9 out of 10 subject methods.
In summary, this paper makes the following contributions:
\begin{itemize}
    \item It proposes \tool, a novel technique to automatically generate valid and effective MRs for functions with numerical, Boolean, array, List and String inputs and outputs.
    \item It describes a framework to automatically evaluate MRs with automated test generation and mutation analysis. 
    \item It presents an empirical evaluation of \tool involving 23 functions from three open-source libraries.
    \item It releases a replication package to facilitate future work~\cite{ourdata}, including the source code of \tool\footnote{\url{https://github.com/jonayerdi/genmorph}} and the results from our experiments.
\end{itemize}

\section{Problem Formulation}
\label{sec:problem}

This section gives the background of this work and formulates the problem of generating effective MRs.
We follow the seminal definition of MRs by Chen et al.~\cite{1998-chen-tr,2017-chen-cs}:

\begin{defin}
Let \emph{f} be a target function or algorithm. A \textbf{Metamorphic Relation (MR)} is a necessary property of \emph{f} over
a sequence of inputs $\langle x_1, \cdots , x_n \rangle$ where $n \geq 2$, and their corresponding outputs $\langle f(x_1), \cdots ,f(x_n)\rangle$.
\end{defin}

In this work, we focus on the subset of MRs that can be represented as a logical implication ($\Rightarrow$) between a relation $R_i$ defined over the sequence of inputs, and a relation $R_o$ defined over the corresponding sequence of outputs. Furthermore, we focus on MRs involving pairs of inputs and corresponding outputs ($n = 2$), which characterize most of the MRs presented in the literature~\cite{2016-segura-tse}. 
\[ R_i(x_1, x_2) \Rightarrow R_o(f(x_1), f(x_2)) \]

Whenever a given input relation $R_i(x_1, x_2)$ holds between the two inputs $x_1$ and $x_2$, a corresponding output
relation $R_o(f(x_1), f(x_2))$ is expected to hold between the outputs.

Referring to the square
function example discussed in the introduction, the presented metamorphic relation is defined as:
\[ (x_1 =  -x_2) \Rightarrow (f(x_1) = f(x_2)) \]

\begin{figure*}[th!]
    \centering
    \includegraphics[width=1\linewidth]{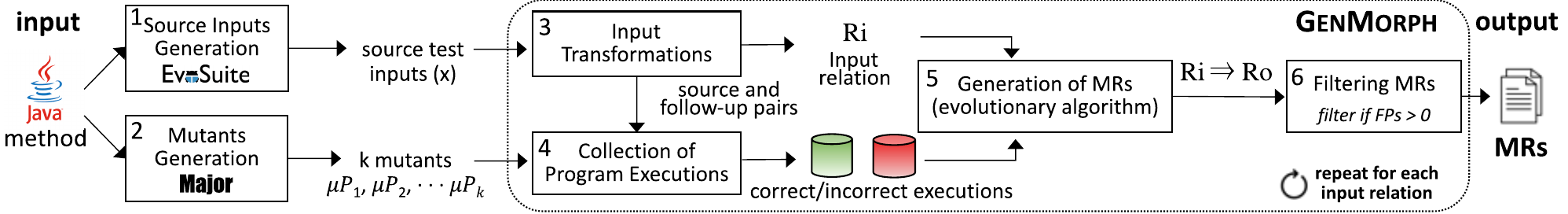}
    \vspace{-6mm}
    \caption{Logical architecture of \tool for the automated generation of MRs}
    \label{fig:approach}
\end{figure*}

A MR can be used as a \textbf{metamorphic test oracle}: an executable Boolean expression that reports an invariant violation if the input relation is satisfied (true), but the output relation is not (false).

Typically, metamorphic test cases are generated by first obtaining an initial test input (by applying any regular test generation strategy), and then applying a transformation to it ($\rightsquigarrow$) such that the input relation will be satisfied. In such cases, the initially generated test input is called the \textbf{source test input} (or source test case), and the one derived from it  to satisfy the input relation is called the \textbf{follow-up test input} (or follow-up test case).
Formally, given an MR, we call a \textbf{(metamorphic) input transformation}~\cite{2018-zhou-tse} a transformation of the inputs $x_1 \rightsquigarrow x_2$ that satisfies $R_i(x_1, x_2)$.

Referring to the square
function  example, an input transformation for its $\textit{MR}$ 
is $x_1 \rightsquigarrow -1 \cdot x_1$. For example, given the source test $x_1 = 4$, its follow-up is $x_2 = -4$.

Given a single MR, MT can generate an arbitrary number of source test inputs and use the \emph{input transformation} to automatically create the corresponding follow-up test inputs. MT executes the function under test on each pair of inputs and reports an oracle violation if the output relation is false. 

Clearly, the effectiveness of metamorphic testing  highly depends on the specific MRs that are used.
Thus, designing effective MRs is a critical step when applying MT~\cite{2016-segura-tse,2017-chen-cs,chen2021new}.
However, discovering effective MRs is labor-intensive and usually requires advanced domain expertise~\cite{1998-chen-tr,2016-segura-tse,2017-chen-cs}. 
%

Given a function \emph{f} and its implementation $P$, \tool aims at automatically generating  one or more MRs that are \textbf{``effective''} at exposing faults in $P$.
We measure effectiveness in terms of false positives (FP) and false negatives (FN) of a test oracle~\cite{oasis,oasistse}.

Jahangirova et al.~\cite{oasis} define a FP as a correct program execution in which the oracle fails but should pass, and a FN as an incorrect program execution in which the oracle passes but should fail.
However, these definitions are ill-suited for MT because metamorphic oracles predicate on multiple test executions.
As such, we modify them as follows:

\begin{defin}
	\label{def:fp}
Let $P$ be an implementation of a target function or algorithm $f$. A \textbf{false positive (FP)} of a metamorphic oracle MR is a pair of inputs $x_1$ and $x_2$ such that both $P(x_1)$ and $P(x_2)$ are correct and \small$[R_i(x_1, x_2) \Rightarrow R_o(P(x_1), P(x_2))] = \textsf{false}$ \normalsize(i.e., \small$R_i(x_1, x_2) = \textsf{true}$ \normalsize and \small$R_o(P(x_1), P(x_2)) = \textsf{false}$\normalsize).
\end{defin}

\begin{defin}
	\label{def:fn}
Let $P$ be an implementation of a target function or algorithm $f$. A \textbf{false negative (FN)} of a metamorphic oracle MR is a tuple $\langle x_1, x_2, \mu P \rangle$, where $x_1$ and $x_2$ is a pair of inputs and $\mu P$ is an incorrect version of $P$ (e.g., seeded fault) such that \small$[ \mu P(x_1) \neq P(x_1)$ \normalsize \textit{or} \small$\mu P(x_2) \neq P(x_2) ]$ \normalsize and \small$[[ R_i(x_1, x_2) \Rightarrow R_o(\mu P(x_1), \mu P(x_2))] = \textsf{true}]$\normalsize.
\end{defin}

In Definition~\ref{def:fp}, 
the condition that both $P(x_1)$ and $P(x_2)$ are correct can be assumed to hold in a regression context, where we are interested in modeling the implemented behavior using a metamorphic oracle that can be adopted later for regression purposes.
In Definition~\ref{def:fn}, 
the condition \small$[ \mu P(x_1) \neq P(x_1)$ \textit{or} $\mu P(x_2) \neq P(x_2) ]$ \normalsize checks that executing the mutant $\mu P$ with at least one of the inputs ($x_1$ or $x_2$) corrupts the output value. If not, the metamorphic oracle has no means to expose the seeded fault.

\begin{probdefinition}{Problem Definition}
\emph{
    Given a function \emph{f} and its implementation $P$, automatically generate one or more MRs that have zero false positives and the fewest false negatives.
}
\end{probdefinition}

There are two important considerations to make:
First, differently from program assertions~\cite{gassert} and pre/post conditions~\cite{evospex}, a single MR usually does not predicate on all possible program inputs and thus can hardly achieve zero FNs.
An MR only predicates on those inputs that satisfy the input relation. 
For this reason, outputting multiple MRs with different input relations is useful as they might complement each other on the types of faults they can detect.
Conversely, aiming at zero FPs is highly desirable because of the high cost of manually inspecting false alarms.

Second, same as Terragni et al.~\cite{gassert} and Molina et al.~\cite{evospex}, we are considering the \emph{implemented} program behavior for false positives, which might differ from the \emph{intended} one. As such, \tool might need a manual validation of the generated MRs to ensure that they capture the intended program behavior.

\section{\tool}

\tool takes as inputs the implementation $P$ of a function (method in Java) and a time budget. It explores the space of possible MRs for $P$, and when the budget expires, it returns the most ``effective'' MRs explored so far.
\tool explores the space of possible MRs with an evolutionary algorithm driven by fitness functions that reward MRs with fewer false positives~(FPs) and false negatives~(FNs).
Figure~\ref{fig:approach} overviews \tool, which generates MRs in six steps:

\smallskip
\noindent \textbf{1) Source Inputs Generation.} 
\tool needs a set of input pairs (source and follow-up tests) to discover and evaluate MRs.
Our current implementation for Java methods employs \textsc{EvoSuite}~\cite{fraser2011evosuite} to generate a diverse set of source test inputs. \textsc{EvoSuite} generates test inputs driven by branch coverage, which guarantees that the generated tests cover diverse execution paths of the method under analysis~\cite{fraser2012whole}.
Since the input relation is still unknown at this stage, \tool uses these generated tests as source test inputs only.
Note that one could also add manually-written test inputs. 

\smallskip
\noindent \textbf{2) Mutants Generation.} 
To obtain a dataset of \emph{incorrect} test executions, \tool executes the generated test inputs on \emph{faulty} versions of the method under test.
To obtain such faulty versions, our current implementation employs \textsc{Major}~\cite{just2014major}, which seeds artificial faults into the method under test, creating a set of $k$ mutants of $P$: $\mu P_{1}, \mu P_{2}, \cdots \mu P_{k}$.
Note that one could also add real faulty versions of $P$ to the set of mutants.

\smallskip
\noindent \textbf{3) Input Transformations.}  \tool generates a set of input relations by relying on predefined templates that specify canonical input transformations~\cite{Duque2022}. These input relations are necessary for generating the follow-up inputs for Step 4.

\smallskip
\noindent \textbf{4) Collection of Program Executions.} 
\tool instruments and executes each pair of source and follow-up inputs $\langle x_1, x_2 \rangle$ to capture at run-time the output values $P(x_1)$, $P(x_2)$ (correct executions) and $\mu P_{i}(x_1), \mu P_{i}(x_2)$ for each mutant $ \forall i~=~1, 2, \cdots k$ (incorrect executions).
Collecting and caching such values is paramount to avoid the cost of re-executing all the tests when computing the FPs and FNs for each explored MR.
Indeed, the evolutionary algorithm of \tool might explore thousands of candidate MRs. 
\tool filters redundant correct and incorrect executions, as well as those incorrect executions that are equivalent to the correct ones obtained with the same input (i.e., it filters $\mu P_{i}(x_1)$ and $\mu P_{i}(x_2)$ if $\mu P_{i}(x_1) = P(x_1)$ and $\mu P_{i}(x_2) = P(x_2)$).

\smallskip
\noindent \textbf{5) Generation of MRs.} 
\tool implements an evolutionary algorithm that, given an input relation and a set of correct/incorrect executions, explores candidate output relations to generate new ones, possibly with zero FP and the fewest FNs. The resulting input and output relations form a complete MR.

\smallskip
\noindent \textbf{6) Filtering MRs.}
Step 5) outputs MRs with zero FPs with respect to the observed correct executions.
Those MRs might fail for some other correct executions. 
To filter such MRs, \tool has a filtering process that uses the test input generator \textsc{Randoop}~\cite{pacheco2007randoop} and the oracle assessor \textsc{OASIs}~\cite{oasis}.

\smallskip
Until the time budget is reached, \tool repeats steps 3) to 6) every time considering a different initial input relation.

\section{Running Example}
\label{sec:example}

This section describes a running example of MR generation using \tool, based on the \emph{pow} function. 
We consider the implementation of the \emph{pow} method from the Apache Commons Math library for Java~\cite{pow} 
(see Listing~\ref{listing:pow}): $\emph{pow}(k, e)$.

\noindent
This method accepts two integers, $k$ and $e$, and returns $k^e$. 

One of the MRs generated by \tool is\footnote{Slightly simplified for clarity and readability}:
\begin{equation} \label{eq:MR_example}
\resizebox{0.91\hsize}{!}{$
    ((k_f = k_s) \wedge (e_f = e_s - 1)) \Rightarrow (pow(k_f, e_f) = \frac{pow(k_s, e_s)}{k_s})
$}
\end{equation}

\noindent
Here, $x_1 = (k_s, e_s)$ denotes the \emph{source} input, and $x_2 = (k_f, e_f)$ the \emph{follow-up} input. In mathematical notation, this MR captures the following property of the \textit{pow} function: $k^e = k^{e-1} \cdot k$.

Listing~\ref{listing:pow} shows a simplified version of the \emph{pow} method. 
For this example, we will use one of the mutants generated by \textsc{Major}, which we will refer to as $\emph{pow}_m$. This mutant removes the \texttt{k2p *= k2p;} statement at line 12, resulting in some incorrect results. The code also shows the instrumentation added by \tool to extract the method inputs and outputs (lines 2 and 15). Listing~\ref{listing:powEvosuite} shows a JUnit test case for $\emph{pow}$ generated by \textsc{EvoSuite}. 
To collect \emph{correct} test executions, \tool executes the JUnit test cases generated by \textsc{EvoSuite} with the instrumented $pow$ method (without mutations).
To collect \emph{incorrect} test executions, \tool
executes the same JUnit test cases with the $\emph{pow}$ mutants generated by \textsc{Major} (which are also instrumented in the same way). Listing~\ref{listing:powState} shows the data generated by the instrumentation after executing the test case in Listing 2 on the instrumented $\emph{pow}$ method. This is the format for correct and incorrect test executions that \tool uses.

\begin{lstlisting}[float, language=Java, label={listing:pow}, caption=Apache Commons Math - pow method (simplified)]
public static int pow(final int k, long e) {
    GenMorph.saveInputValues(k, e) // instrumentation
    if (e < 0) {
        throw new NotPositiveException();
    }
    int result = 1;
    int k2p = k;
    while (e != 0) {
        if ((e & 0x1) != 0) {
            result *= k2p;
        }
        k2p *= k2p; // Mutation: Remove line
        e >>= 1;
    }
    GenMorph.saveOutputValues(result) //  instrumentation
    return result;
}
\end{lstlisting}

\begin{lstlisting}[float, language=Java, label={listing:powEvosuite}, caption=Test input for pow generated by \textsc{EvoSuite},  belowskip=-0.4 \baselineskip]
@Test(timeout = 4000)
public void test100()  throws Throwable  {
  int int0 = ArithmeticUtils.pow(-128, 2);
}

\end{lstlisting}

\begin{lstlisting}[float, language=Python, label={listing:powState}, caption=\small GenMorph (correct) execution by executing Listing~\ref{listing:powEvosuite}]
    "systemId": "pow@original", "testId": "test100",
    "variables": { "inputs": { "k": -128.0, "e": 2.0},
                   "outputs": {"return": 16384.0} }
\end{lstlisting}
Let us consider the input relation ($(k_f = k_s) \wedge (e_f = e_s - 1)$).
Given the source input ($k=-128$, $e=2$) (Listing~\ref{listing:powEvosuite}), the follow-up input is ($k_f=-128$, $e_f=1$). Executing the original $\emph{pow}$ (Listing~\ref{listing:pow}) with these source and follow-up inputs yields respectively the following (correct) outputs:
\[
pow(-128, 2) = 16384 \quad 
pow(-128, 1) = -128 
\]
Executing the faulty version generated with \textsc{Major} ($\emph{pow}_m$) with the same inputs yields the following outputs:
\[
pow_m(-128, 2) = -128 \quad 
pow_m(-128, 1) = -128 
\]

If we consider these correct and incorrect executions, \tool infers that the generated MR shown in Eq.\ref{eq:MR_example} correctly classifies these executions. This MR identifies $\emph{pow}_m$ as faulty since the input relation is satisfied, but the output relation is violated. At the same time, the original $\emph{pow}$ satisfies the MR.

Obviously, there are too few test executions used in this example to capture the behavior of $\emph{pow}$, so other (invalid) MRs could also be inferred from the considered executions, such as:
\begin{equation} \label{eq:MR_example_bad}
  \resizebox{0.91\hsize}{!}{ $((k_f = k_s) \wedge (e_f = e_s - 1)) \Rightarrow (pow(k_f, e_f) \neq pow(k_s, e_s))$}
\end{equation}
This MR is incorrect because it yields FPs when $k=0$ or $k=1$.  
Consider the following input and output pair, generated with the same mutant:
$pow_m(2, 8) = 2 \quad 
pow_m(2, 7) = 8$.
The invalid MR (Equation \ref{eq:MR_example_bad}) would pass, whereas our example MR (Equation \ref{eq:MR_example}) would correctly identify this faulty execution.

This example shows that intermediate MRs might have both FPs and FNs. That is why \tool uses genetic programming (GP) to improve candidate MRs until they have no FPs and fewer FNs.

\section{Input Transformations}
\tool starts the generation of MRs from the input relations. 
From a set of predefined transformation templates, \tool selects a series of (metamorphic) input transformations compatible with the given method under test. 
For each selected transformation and source test input, \tool generates a follow-up test input. 
Furthermore, it also generates a \emph{canonical input relation} from each selected transformation.
Such input relation is the most strict interpretation of the applied transformation: Given the values of the source test inputs, there is only a single possible value of the follow-up inputs that satisfies the canonical input relation.

These transformation templates are similar to what the MR literature refers to as metamorphic relation input patterns (MRIPs): An abstraction that characterizes the relations among the source and follow-up inputs of a set of MRs \cite{2018-zhou-tse}. Our templates, however, define not only the input relations, but also the corresponding transformations from the source to the follow-up inputs.

Currently, \tool implements the following \textbf{transformation templates}. 
We derived such templates by referring to the recent study of Duque et al.~\cite{Duque2022} that reports MRs commonly used in the literature (e.g.,~\cite{kanewala2013using,kanewala2014techniques}).

\smallskip
\noindent \textbf{PermuteParameters}.
It permutes two of the input parameters of the method. It can be applied to any pair of method parameters, as long as they have the same type. For example, there is one possible instantiation for $\emph{pow}$, in which the follow-up for $\emph{pow}(k, e)$ is $\emph{pow}(e, k)$. In this example, the canonical input relation would be:
    $(k_f = e_s) \wedge (e_f = k_s)$.
    
\smallskip
\noindent \textbf{BooleanFlip}. It flips the value of a Boolean parameter from \emph{true} to \emph{false} or vice versa. This transformation cannot be applied to $\emph{pow}$ because both of its parameters are numeric.
Although the study of Duque et al.~\cite{Duque2022} does not mention this relation, we add it to support input relations of Boolean parameters.
    
\smallskip
\noindent \textbf{NumericAddition(Number)}. It adds a constant number (positive or negative) to a single numeric parameter.  
For $\emph{pow}$, we could apply \texttt{NumericAddition(-1)} to its second parameter to generate the follow-up input $pow(k, e - 1)$.     This is the transformation template used by \tool to generate the MR discussed in Section~\ref{sec:example}.
The corresponding canonical input relation would be: $(k_f = k_s) \wedge (e_f = e_s - 1)$.

\smallskip
\noindent \textbf{NumericMultiplication(Number)}. This transformation multiplies a single numeric parameter by a constant number (positive or negative).  
For $\emph{pow}$, we can apply \texttt{NumericMultiplication(2)} to its first parameter and generate the follow-up input $\emph{pow}(k \cdot 2, e)$. The corresponding input relation would be: $(k_f = k_s \cdot 2) \wedge (e_f = e_s)$.

\smallskip
\noindent \textbf{SequenceRemove(Number)}. It removes a single element from a Sequence parameter. The numeric parameter of this template indicates the index to be removed, with negative indices being allowed for backwards indexing (like Python list indexing). If the index is out of bounds, this operation is a no-op. This transformation cannot be applied to $\emph{pow}$ because both of its parameters are numeric.

\smallskip
\noindent \textbf{SequenceFlip}. It inverts the order of the elements from a Sequence parameter. This transformation cannot be applied to $\emph{pow}$ because both of its parameters are numeric.
Although the study of Duque et al.~\cite{Duque2022} does not mention this relation directly, it is conceptually similar to permuting parameters, and has been applied in various MRs throughout the literature.

\smallskip

Except for trivial functions (e.g., those that have a single Boolean parameter), there are many possible applicable input transformations to a given function (using the above templates). 
Indeed, the parameterized transformation templates (i.e., \emph{NumericAddition} and \emph{NumericMultiplication}) have a myriad of possible instantiations, as their parameter can be any number.

Because of this, it is necessary to sample a meaningful set of values for the parameterized transformation templates. 
To this aim, \tool collects a pool of constants, consisting of the predefined values -1 and  1, plus all the constant values appearing in the method under test. \tool extracts the constant values by instrumenting all the variable accesses and literal expressions of the method under test. Then it executes all the source inputs generated by \textsc{EvoSuite} in Step 1, and identifies all the values logged by the instrumentation that remain the same in all executions. Such constants are likely to represent meaningful values for the method under test that might lead to semantically meaningful MRs.  \tool randomly samples the constant values to use with parameterized transformations (prioritizing constants that appeared more often).

\section{Generation of MRs}
\tool explores the space of possible MRs to find one that accurately classifies correct and incorrect test executions. 
Unfortunately, the space of possible MRs is enormous. 
As such, \tool employs GP~\cite{koza1992genetic} to guide the search.

Similarly to \gassert~\cite{gassert}, \tool implements a co-evolutio\-na\-ry algorithm that evolves two populations of MRs in parallel, with three competing objectives:
\begin{inparaenum}[(i)]
	\item minimizing the FPs,
	\item minimizing the FNs,
	\item minimizing the size of the MR.
\end{inparaenum}
Each population uses a different fitness function.
The fitness function used in the first population ($\phi_{\textit{FP}}$) rewards MRs with fewer FPs, while the function of the second population ($\phi_{\textit{FN}}$) those with fewer FNs.
Both populations consider the remaining two objectives only in tie cases.
Periodically, the two populations exchange their best individuals to provide good genetic material to improve the secondary objectives.

\smallskip
\textbf{Fitness Functions.} Let $\textit{FP}(\textit{MR})$ and $\textit{FN}(\textit{MR})$ denote the false positive and false negative rate of an individual $\textit{MR}$  (with respect to the given correct and incorrect executions), and let $|MR|$ be the size of the MR (i.e., the number of syntactic elements in its predicate).
The multi-objective fitness functions $\phi_{\textit{FP}}$ and $\phi_{\textit{FN}}$~\cite{gassert} are defined using the concept of \textit{dominance}~($\prec$)~~\cite{deb:nsgaii:tec:2002}:
\begin{defin}
	\textbf{FP-fitness ($\mathbf{\phi_{\textit{FP}}}$)}. Given two metamorphic relations $MR_1$ and $MR_2$, $MR_1$ \textbf{dominates}$_{\mathbf{FP}}$ $MR_2$ ($MR_1$ \textbf{$\mathbf{\prec}_{\textit{FP}}$} $MR_2$)   if any of the following conditions is satisfied:

	\footnotesize
	\begin{itemize}
		\item[--] $\textit{FP}(MR_1) < \textit{FP}(MR_2)$
		\item[--] $\textit{FP}(MR_1) = \textit{FP}(MR_2), \textit{FN}(MR_1) < \textit{FN}(MR_2)$
		\item[--] $\textit{FP}(MR_1) = \textit{FP}(MR_2),  \textit{FN}(MR_1) = \textit{FN}(MR_2), \mid$$MR_1$$\mid < \mid$$MR_2$$\mid$
	\end{itemize}
\end{defin}

\begin{defin}
	\textbf{FN-fitness ($\mathbf{\phi_{\textit{FN}}}$)}. Given two metamorphic relations $MR_1$ and $MR_2$, $MR_1$ \textbf{dominates}$_{\mathbf{FN}}$ $MR_2$ ($MR_1$ \textbf{$\mathbf{\prec}_{\textit{FN}}$} $MR_2$)  if any of the following conditions is satisfied:
	
	\footnotesize
	\begin{itemize}
		\item[--] $\textit{FN}(MR_1) < \textit{FN}(MR_2)$
		\item[--] $\textit{FN}(MR_1) = \textit{FN}(MR_2), \textit{FP}(MR_1) < \textit{FP}(MR_2)$
		\item[--] $\textit{FN}(MR_1) = \textit{FN}(MR_2),  \textit{FP}(MR_1) = \textit{FP}(MR_2), \mid$$MR_1$$\mid < \mid$$MR_2$$\mid$
	\end{itemize}
\end{defin}

In tie cases, {\small $\textit{FP}(MR_1)$$=$$\textit{FP}(MR_2)$} and {\small $\textit{FN}(MR_1)$$=$$\textit{FN}(MR_2)$}, $\phi_{\textit{FP}}$ and $\phi_{\textit{FN}}$ favor smaller MRs, which are easier to understand.

Function \textbf{\textsc{MR-generation}} of Algorithm~\ref{algo:evo} describes our co-evolutionary approach. First, \tool initializes two distinct populations (\textit{Popul}$^{\textit{FP}}$ and \textit{Popul}$^{\textit{FN}}$) with MRs composed of the given input relation ($R_i$) and randomly generated output relations ($R_o$) (lines~\ref{a:initpopfp} and~\ref{a:initpopfn}). Then, until the time budget is expired, \tool evolves each population in parallel 
(Function~\textsc{select+reproduce}). \textit{Popul}$^{\textit{FP}}$ uses $\phi_{\textit{FP}}$ to select the individuals, while \textit{Popul}$^{\textit{FN}}$ uses $\phi_{\textit{FN}}$.
Periodically, the two populations exchange their best individuals (lines~\ref{a:migrationstart} to \ref{a:migrationend}).

	\setlength{\algomargin}{4mm}
	\begin{algorithm}[t!]
	\DontPrintSemicolon
	 \SetKwBlock{DoParallel}{do in parallel}{end}
	\linespread{0.8}\selectfont
	\footnotesize
	\SetAlCapHSkip{0em}
	\SetKwProg{Fn}{function}{}{}
	\SetKwInOut{Input}{input}
	\SetKwInOut{Output}{output}

\Input{
$R_i$: canonical input relation \\
correct $\mathcal{E}^+$ and incorrect $\mathcal{E}^-$ test executions}

\Output{$\mathbb{MR}$ a set of the best metamorphic relations}
\BlankLine
\Fn{\textsc{\textbf{MR-generation}}}{

	    \textit{Popul}$^{\textit{FP}}$ $\leftarrow$ \textsc{get-initial-random-population}($R_i$) \label{a:initpopfp}\;
		\textit{Popul}$^{\textit{FN}}$ $\leftarrow$ \textsc{get-initial-random-population}($R_i$)\label{a:initpopfn}\;
		
		\textit{gen} $\leftarrow 0$\;
		
		\Repeat{ time budget is expired}{
		\textit{gen} $\leftarrow \textit{gen} + 1$\;

        \DoParallel{
        	\textit{Popul}$^{\textit{FP}}$\hspace{-0.5mm}$\leftarrow$\textsc{\textbf{select+reproduce}}(\textit{Popul}$^{\textit{FP}}$, $\phi_{\textit{FP}}$, \textit{gen})\label{a:evofp}\;
        	\textit{Popul}$^{\textit{FN}}$\hspace{-0.5mm}$\leftarrow$\textsc{\textbf{select+reproduce}}(\textit{Popul}$^{\textit{FN}}$, $\phi_{\textit{FN}}$, \textit{gen})\label{a:evofn}\;
        }
    
   \If{gen \% FREQ\_MIGRATION = 0\label{a:migrationstart}}{
    	add \textsc{get-best-MRs}(\textit{Popul}$^{\textit{FN}}$, $\phi_{\textit{FN}}$) to \textit{Popul}$^{\textit{FP}}$\;
    	add \textsc{get-best-MRs}(\textit{Popul}$^{\textit{FP}}$, $\phi_{\textit{FP}}$) to \textit{Popul}$^{\textit{FN}}$\label{a:migrationend}\;
    }
}
\Return{\textsc{get-best-MRs }(\{\textit{Popul}$^{\textit{FP}} \cup  \textit{Popul}^{\textit{FN}}\}$, $\phi_{\textit{FP}}$)} 
}
\BlankLine
\Fn{\textsc{\textbf{select-+-reproduce}}}{
\textit{Popul} $\leftarrow$ \textsc{compute-fitness}(\textit{Popul}, $\phi$, $\mathcal{E}^+$, $\mathcal{E}^-$)\label{a:compute}\;
		\textit{Popul}$_{\textsc{new}}$ $\leftarrow$ \textsc{get-best-MRs }(\textit{Popul}, $\phi$)\label{a:initinew}\tcp*{\small elitism}
		
	\Repeat{Popul$_{\textsc{new}}$ is full \label{a:whilepop}}{
	    $\langle \textit{MR}_{p1}, \textit{MR}_{p2} \rangle \leftarrow $\textsc{select-parents}(\textit{Popul}, $\phi$)\label{a:select}\;

         $\langle \textit{MR}_{o1},\textit{MR}_{o2}$$\rangle$$\leftarrow $\textsc{crossover+mutation}($\textit{MR}_{p1}$$,\textit{MR}_{p2})$\label{a:crossover}
		add $\langle \textit{MR}_{o1}, \textit{MR}_{o2} \rangle$ to \textit{Popul}$_{\textsc{new}}$\label{a:addoff}\;
      }
	\Return{\textit{Popul}$_{\textsc{new}}$\label{a:returnpop}}
}

	\caption{\label{algo:evo} \\ \textsc{\tool Co-Evolutionary Algorithm}}
\end{algorithm}

\subsection{Selection, Crossover, and Mutation}

Function \textbf{\textsc{select+reproduce}} of Algorithm~\ref{algo:evo} describes how each generation of \tool evolves a population of MRs ($Popul$ into $Popul_{\textsc{new}}$). First, it computes the fitness of each (new) individual in the population (Line~\ref{a:compute}). This amounts to counting the number of FPs, FNs, and computing the size of the MRs. Second, \tool initializes the new population $Popul_{\textsc{new}}$ with the best individuals (elitism) to ensure that they will not be lost (Line~\ref{a:initinew}).
Then, Function \textsc{select+reproduce} applies Selection, Crossover, and Mutation. The \textbf{Selection} step selects two parent individuals $\langle \textit{MR}_{p1}, \textit{MR}_{p2} \rangle $ based on the given fitness function, $\phi_{\textit{FN}}$ for \textit{Popul}$^{\textit{FN}}$ and $\phi_{\textit{FP}}$ for \textit{Popul}$^{\textit{FP}}$ (Line~\ref{a:select}).
 The \textbf{Crossover} step combines the genetic materials (portions of MRs in our case) of the selected individuals and produces two new MRs (offspring) $\langle \textit{MR}_{o1}, \textit{MR}_{o2} \rangle$ (Line~\ref{a:crossover}). Finally, the \textbf{Mutation} step mutates (with a certain probability) the obtained offspring (Line~\ref{a:crossover}) and adds them to the new population. It repeats these three steps until the new population is full. 

\subsubsection{Selection}
 \tool implements two different selection criteria, and chooses between them with a given probability.

\emph{Tournament Selection}~\cite{miller1995genetic} runs two ``tournaments'' among $K$ random individuals.
The winner of each tournament (the one with the highest fitness) is selected as a parent.
We chose $K = 2$, as it mitigates the \textit{local optima problem}~\cite{whitley:GA:1994}.

\emph{Best-match Selection}~\cite{gassert} is a selection criterion specific for test oracles.
It selects the first parent randomly and selects the second parent with a higher probability if it maximizes the collective number of covered correct and incorrect executions.

\subsubsection{Crossover and Mutation}

\tool represents output relations as rooted binary trees, where  nodes can be operators, constant values, or variables.
Table~\ref{table:GenMorphOperators} shows the operators supported by \tool. All numeric values are treated as real numbers (with floating point, fixed precision representation), and the numeric constants can have any real value in the range $[-100, 100]$. On the other hand, $true$ and $false$ are Boolean constants, and there are no sequence-type constants. As for the variables, the output relation can contain any input or output variable from the source or the follow-up inputs.

\smallskip 
\noindent
\textbf{Crossover}.
\tool relies on the classical tree-based crossover~\cite{koza1992genetic}. It selects a random crossover point in the output relations of each parent and creates two trees by swapping the subtrees rooted at each point.

\smallskip 
\noindent
\textbf{Mutation}.
\tool relies on three tree-based mutation operators (chosen randomly with a given prob.).

\emph{Node Mutation} changes a single node in the tree~\cite{brameier2007comparison}.
Given a tree and one of its leaf nodes $n$, the new tree has $n$ replaced with a new node of the same type (generated randomly).

\emph{Subtree Mutation} replaces a subtree in the tree~\cite{brameier2007comparison}.
Given a tree and one of its inner nodes $n$, it returns a new tree obtained by substituting the subtree rooted at $n$ with a randomly generated subtree of the same type.

\begin{table}[t]
\centering
\renewcommand*{\arraystretch}{0.80}
\setlength{\tabcolsep}{10pt}
\caption{Operators considered by \tool}
\vspace{-3mm}
\label{table:functions}
	\resizebox{\linewidth}{!}{%
				\rowcolors{1}{}{gray!10}
\begin{tabular}{lll}
				\hiderowcolors

\toprule
\textbf{Operands} & \textbf{Output} & \textbf{Function} \\
\midrule
\showrowcolors
$\langle number, number \rangle$ & $number$ & $+$, $-$, $*$, $/$ (protected division) \\
$\langle number \rangle$ & $number$ & ABS \\
$\langle number, number \rangle$ & $Boolean$ & $==$, $\neq$, $<$, $>$, $\leq$, $\geq$ \\
$\langle number \rangle$ & $sequence$ & toString \\
$\langle Boolean, Boolean \rangle$ & $Boolean$ & AND, OR, XOR, iff, implies \\
$\langle Boolean \rangle$ & $Boolean$ & NOT \\
$\langle sequence \rangle$ & $number$ & length, sum \\
$\langle sequence \rangle$ & $boolean$ & $==$, $\neq$ \\
$\langle sequence \rangle$ & $sequence$ & flip \\
$\langle sequence, number \rangle$ & $sequence$ & remove, truncate \\
\bottomrule
\end{tabular}
\label{table:GenMorphOperators}
}
\end{table}

\emph{Constant Value Mutation} changes the value of a numeric constant node.
It takes a tree as its only input, and returns a new tree obtained by randomly selecting a numeric constant node and adding a random number, chosen from $\{-\Delta, \Delta\}$.
Here, $\Delta$ should be small (we use $0.1$ in our experiments) so that the constant values change in small increments.

\subsection{Constraints} 

\tool constrains the generated MRs. If an individual violates one or more of these constraints, it is dropped immediately, without  evaluating its fitness. Currently, the output relations have a configurable complexity limit, which is the maximum number of nodes that its tree can have, and any individual which surpasses this limit will not be added to the population.

Furthermore, we also implemented a ``soft'' constraint, which is required for an individual to be considered for the elite set or as the best individual, but not for being added to the population.
This constraint is that the output relation must contain both the source and the follow-up variables for at least one of the method outputs. This filters out expressions that are not truly MRs, as MRs are supposed to check the outputs from both the source and the follow-up test cases. Although individuals that violate this constraint are not MRs, they are allowed into the population because they may contain useful genetic material.

Finally, we also added uniqueness constraints for elitism. Specifically, an individual must satisfy two requirements to be included in the elite population, on top of its fitness. First, there must not be another individual with an identical output relation already in the population. Second, there must not be another individual with an identical set of FNs already in the population. These constraints prevent semantically equivalent individuals from taking multiple spots in the elite population. Notably, the second constraint is needed because simple mutations can easily bypass the first constraint and generate semantically equivalent individuals. For instance: $ASSERTION$ may become $(ASSERTION \land true)$.

\section{MR filtering}
\label{sec:filtering}

\tool performs a final filtering process to avoid reporting \emph{invalid} MRs to the user. The validity of an MR is determined by the lack of FPs after a 2-step filtering process. 

The first step validates the MRs with new automatically-generated test inputs unseen by the evolutionary algorithm.
We use the random test generator \textsc{Randoop}~\cite{pacheco2007randoop} to generate the new source inputs.
Then, \tool can simply generate the corresponding follow-up inputs using the template-based input transformation corresponding with the input relation of each MR.

For each MR under analysis, \tool creates a JUnit test suite with a test case for each source and follow-up input pair generated with \textsc{Randoop} and the input transformation for the MR. Listing \ref{listing:junit} shows a JUnit test case generated for the example MR shown in Eq.~\ref{eq:MR_example}. 
Each failing test would be a FP for the analyzed MR.

The second step relies on OASIs~\cite{oasis,oasistool,oasistse}, an oracle assessor designed to detect FPs and FNs in program assertions. 
To detect FPs, it negates the assertion and creates a new branch at the assertion point with the negated condition. It then employs search-based test generation to find test cases that cover that branch, which would represent FPs of the assertion.

To adapt OASIs' functionality to MRs, we changed its original implementation.
Given an MR, OASIs creates a new method that, similarly to Listing~\ref{listing:junit}, has an if statement with the input relation as its condition and an assert statement with the output relation as its body.
However, unlike Listing~\ref{listing:junit}, the generated method is not input specific, but takes the source and follow-up inputs as parameters. 
This new branch becomes the target of the search-based test generation of OASIs.  
This enables the detection of FPs, as the inputs that make the assert statement fail will need to satisfy the input relation first.

Note that the filtering step ignores the FNs of MRs. As discussed in Section~\ref{sec:problem}, FNs are inevitable in MRs, while ensuring absence of FPs is crucial to avoid  false alarms.

\begin{lstlisting}[float, language=Java, label={listing:junit}, caption= \tool-generated executable MR of Eq. (1)]
@Test
public void test0followup() {
    int k_s = -128, e_s = 2; // source input
    int k_f = -128, e_f = 1; // follow-up input
    int o_s = pow(k_s, e_s); // run source test input
    int o_f = pow(k_f, e_f); // run follow-up test input
    if ((k_f == k_s) && (e_f == e_s - 1)) {//Ri is true
        assertTrue(o_f == (o_s / k_s)); // check Ro 
    }
}
\end{lstlisting}

\section{Evaluation} \label{sec:evaluation}
\begin{table*}[h]
\setlength{\tabcolsep}{2.6pt}
\renewcommand{\arraystretch}{0.95}
\caption{Evaluation results (average) (MS = Mutation Score, PZ = Ratio of MRs without FPs, PZO = Ratio of MRs without FPs with OASIs, $\Delta$MS = \tool's MS over mutants missed by \textsc{Randoop}~\cite{pacheco2007randoop} ($\Delta$MS$_R$), \textsc{Evosuite}~\cite{fraser2011evosuite} ($\Delta$MS$_E$) or \textsc{AutoMR}~\cite{automr} ($\Delta$MS$_A$)).}
\rowcolors{1}{}{gray!10}
\begin{tabular}{lllr|r|r|r|rrrrrr}
\toprule
\hiderowcolors
\multirow{2}{*}{library} & \multirow{2}{*}{method} & \multirow{2}{*}{signature} & \multirow{2}{*}{\# mutants} & \multicolumn{1}{c|}{Randoop} & \multicolumn{1}{c|}{Evosuite} & \multicolumn{1}{c|}{AutoMR} & \multicolumn{5}{c}{\tool} \\
&  &  &  & \multicolumn{1}{c|}{MS} & \multicolumn{1}{c|}{MS} & \multicolumn{1}{c|}{MS} & \multicolumn{1}{c}{MS} & $\Delta$MS$_R$ & $\Delta$MS$_E$ & $\Delta$MS$_A$ & \multicolumn{1}{c}{PZ} & \multicolumn{1}{c}{PZO} \\
\midrule
\showrowcolors
Math & nextPrime & \texttt{int(int)} & 20 & 0.80 & 0.90 & 0.28 & 0.78 & 0.75 & 0.46 & 0.72 & 0.70 & 0.97 \\
Math & isPrime & \texttt{bool(int)} & 25 & 0.52 & 0.92 & 0.28 & 0.29 & 0.67 & 0.06 & 0.17 & 0.32 & 1.00 \\
Math & gcd & \texttt{int(int,int)} & 25 & 0.50 & 0.83 & 0.00 & 0.63 & 0.59 & 0.05 & 0.63 & 0.90 & 1.00 \\
Math & pow & \texttt{int(int,int)} & 10 & 0.70 & 1.00 & 0.00 & 0.69 & 0.00 & - & 0.69 & 0.78 & 1.00 \\
Math & stirling & \texttt{long(int,int)} & 46 & 0.28 & 0.56 & 0.00 & 0.31 & 0.24 & 0.20 & 0.31 & 0.69 & 0.79 \\
Math & acos & \texttt{double(double)} & 76 & 0.93 & 0.50 & 0.00 & 0.09 & 0.02 & 0.08 & 0.09 & 0.05 & 0.71 \\
Math & log10 & \texttt{double(double)} & 15 & 1.00 & 0.60 & 0.00 & 0.06 & - & 0.00 & 0.06 & 0.03 & 0.32 \\
Math & sin & \texttt{double(double)} & 26 & 0.71 & 0.73 & 0.41 & 0.60 & 0.00 & 0.00 & 0.36 & 0.70 & 0.97 \\
Math & sinh & \texttt{double(double)} & 123 & 0.89 & 0.41 & 0.23 & 0.21 & 0.00 & 0.10 & 0.20 & 0.25 & 0.55 \\
Math & tan & \texttt{double(double)} & 37 & 0.76 & 0.70 & 0.32 & 0.38 & 0.00 & 0.01 & 0.25 & 0.56 & 0.79 \\
Lang & abbreviate & \texttt{string(string,string,int,int)} & 39 & 0.85 & 0.30 & - & 0.41 & 0.01 & 0.13 & - & 0.53 & 0.96 \\
Lang & capitalize & \texttt{string(string)} & 10 & 0.90 & 0.41 & - & 0.28 & 0.00 & 0.00 & - & 0.13 & 1.00 \\
Lang & center & \texttt{string(string,int,string)} & 12 & 0.83 & 0.25 & - & 0.53 & 0.00 & 0.28 & - & 0.64 & 0.94 \\
Lang & difference & \texttt{string(string,string)} & 6 & 0.67 & 0.36 & - & 0.24 & 0.00 & 0.50 & - & 0.19 & 0.90 \\
Lang & isSorted & \texttt{bool(int[])} & 11 & 0.90 & 0.37 & - & 0.18 & 0.25 & 0.25 & - & 0.11 & 0.80 \\
Guava & indexOf & \texttt{int(bool[],bool[])} & 12 & 1.00 & 1.00 & - & 0.14 & - & - & - & 0.04 & 0.40 \\
Guava & join & \texttt{string(string,bool[])} & 5 & 0.80 & 0.80 & - & 0.13 & 0.00 & 0.00 & - & 0.06 & 1.00 \\
Guava & meanOf & \texttt{double(int[])} & 12 & 0.96 & 1.00 & - & 0.83 & 0.00 & - & - & 0.88 & 1.00 \\
Guava & min & \texttt{int(int[])} & 9 & 0.89 & 0.89 & - & 0.55 & 0.00 & 0.00 & - & 0.47 & 0.74 \\
Guava & padStart & \texttt{string(string,int,char)} & 7 & 0.86 & 0.83 & - & 0.64 & 0.00 & 0.16 & - & 0.35 & 0.85 \\
Guava & repeat & \texttt{string(string,int)} & 18 & 0.94 & 0.81 & - & 0.86 & 0.00 & 0.69 & - & 0.67 & 0.97 \\
Guava & sort & \texttt{void(byte[],int,int)} & 8 & 0.95 & 0.88 & - & 0.58 & 0.00 & 0.00 & - & 0.51 & 0.83 \\
Guava & truncate & \texttt{string(string,int,string)} & 10 & 1.00 & 0.78 & - & 0.28 & - & 0.18 & - & 0.23 & 0.80 \\
\bottomrule
\end{tabular}
\label{table:aggregate}
\end{table*}

Our evaluation aims to answer three research questions (RQs).

\begin{itemize}
    \item[\textbf{RQ1: Effectiveness}] Is \tool effective at synthesizing valid and useful metamorphic relations automatically?
    
    \item[\textbf{RQ2: Test Case Oracle Enhancement}] How much does \tool increase the fault detection capability of automatically generated test inputs and test case oracles?

    \item[\textbf{RQ3: AutoMR Comparison}] How do the results from \tool compare with those obtained by \textsc{AutoMR} \cite{automr}?

    \item[\textbf{RQ4: Filtering}] Is the filtering process of \tool effective at detecting invalid metamorphic relations?
\end{itemize}

RQ1 focuses on the main objective of \tool, which is the generation of oracles that do not trigger false alarms on the original code (\textit{valid} oracles), while being effective at exposing faults injected by mutation analysis (\textit{useful} oracles).
To be effective, metamorphic oracles should also pass the OASIs filter, which excludes metamorphic oracles for which evidence of false alarms can be automatically generated.
RQ2 compares the fault detection capability of \tool with that of test generators that produce test cases with assertions on the observed execution output for specific test inputs (henceforth referred to as \textit{test case oracles}). Similarly to the regression oracles generated by \tool, which capture universal properties, these test case oracles can be useful to expose faults in a regression testing context. RQ3 compares \tool with \textsc{AutoMR} \cite{automr}, a state-of-the-art MR generation tool that can generate polynomial MRs for programs with numeric inputs and outputs. RQ4 focuses on the usefulness of the the proposed filtering process, which uses automated test generation to find evidence of false positives and to filter out the corresponding metamorphic oracles.

\subsection{Experimental Setup}

\noindent \textbf{Experimental Subjects.}
We evaluated \tool on three different open-source Java libraries: (1) Apache Commons Math 3~\cite{math}, (2) Apache Commons Lang 3~\cite{lang}, and (3) Google Guava 31~\cite{guava}.
We randomly selected the methods from the ones in the libraries that fulfill the following criteria: (1) it contains at least 8 lines of code. The methods we selected contain between 8 and 93 LOC; (2) it is a static method (since \tool does not currently consider the \texttt{this} object state); (3) all the method parameters are currently supported by \tool. The supported types include Java primitives (e.g., \texttt{boolean}, \texttt{int}) and their corresponding wrappers (e.g., \texttt{java.lang.Boolean}, \texttt{java.lang.Integer}), or types which we consider sequences in our implementation (arrays, \texttt{java.lang.List} implementations, or \texttt{java.lang.String});
(4) it does not use I/O or non-deterministic operations (e.g., reading/writing files, using random numbers).
The 23 selected methods span 13 different classes from the three different libraries.
The first four columns of Table~\ref{table:aggregate} show their library, name, signature, LOC, \# of mutants.

\smallskip
\noindent \textbf{Setup.}
We now describe the evaluation setup for \tool.

\emph{\textbf{Step 1)} Source Input Generation.}
For each subject, we ran \textsc{EvoSuite}~\cite{fraser2011evosuite} (v. 1.1.0) to generate the source test inputs.
We configured \textsc{EvoSuite} with the branch coverage criterion, minimization enabled, and a time budget of 5 minutes. We performed ten runs with
different random seeds and aggregated all the test cases to obtain a diverse and large set of test inputs.

\emph{\textbf{Step 2)} Mutants Generation.}
For each subject, we ran \textsc{Major}~\cite{just2014major} (v. 2.0.0) enabling all types of supported mutants.
We automatically filtered the mutants by mutated lines of code, keeping only those that directly affect the
method under test.

\emph{\textbf{Step 3)} Input Transformation.}
In order to keep the MR generation at a reasonable cost, we configure \tool to select four instantiations of our templates for each run. 

 \emph{\textbf{Step 4)} Collection of Program Executions.}
Since the cost of evaluating the fitness of an MR grows linearly with the number of correct and incorrect executions, we enforced a limit of 9,000 unique correct and 9,000 unique incorrect executions. If more are obtained, the executions are randomly sampled.

 \emph{\textbf{Step 5)} Generation of MRs.}
Table \ref{table:Configuration} shows the configuration of Algorithm~\ref{algo:evo}.

\begin{table}[b]
\vspace{-2mm}
	\setlength{\tabcolsep}{2pt}
	\renewcommand{\arraystretch}{0.85}
	\vspace{-3mm}
\caption{Configuration Parameters of Algorithm~\ref{algo:evo}}
	\resizebox{1\linewidth}{!}{%
		\rowcolors{1}{}{gray!10}
\begin{tabular}{lr|lr}
\hiderowcolors
\toprule
\textbf{Parameter Description} & \textbf{Value} & \textbf{Parameter Description} & \textbf{Value} \\ \midrule
\showrowcolors
time budget (minutes) & $30$ & prob. of crossover & $90\%$ \\
max correct executions & 9,000 & prob. of mutation & $30\%$ \\
max incorrect executions & 9,000 & frequency of migration (every X gen) & $10$ \\
bound on the size of $R_o$ & $16$ & number of individuals for elitism & $10$ \\
size of each of the populations & 1,000 & number of individuals to migrate & $160$ \\
\bottomrule
\end{tabular}
}
\label{table:Configuration}
\end{table}
For each subject method and strategy, we used a time budget of 30 minutes and we generated four different MRs.
We repeated each run 12 times due to the stochasticity of the approaches.

\emph{\textbf{Step 6)} Filtering MRs.}
For each subject, we ran \textsc{Randoop}~\cite{pacheco2007randoop} (v. 4.3.0)  with 12 different random seeds. In order to limit the run-time of the experiments to a reasonable cost, we sampled 100 \textsc{Randoop} test cases for each run.
We ran OASIs~\cite{oasis} multiple times (due to its stochastic nature) with an overall budget of 10 minutes, giving each run a budget of 150 seconds. We stop OASIs when we find the first FP.

We ran the experiments on four VMs with a 6-core AMD Zen 3.2 GHz CPU and 12 GB of RAM.

\subsection{RQ1: Effectiveness}
RQ1 evaluates the effectiveness of \tool at synthesizing effective MRs using Mutation Score (MS) (i.e., the ratio of killed mutants).
For each MR returned by Step 5, we ran the 2-step filtering process (\textsc{Randoop}, and \textsc{OASIs}, see Section~\ref{sec:filtering}). 
If the filtering process did not find FPs, we passed each of the 12 test suites derived from the \textsc{Randoop} test inputs and the generated MR to PIT~\cite{coles2016pit} (v. 1.7.4) for computing the MS. 
This results in 12 different MS measures for each MR. If the 2-step filtering process found FPs, we considered the MS to be zero.
Note that during MR generation, \tool relies on completely separate tools (i.e., \textsc{EvoSuite} and \textsc{Major}) to obtain the incorrect executions. In the empirical study, using different test inputs and mutants to compute the MS is important to properly separate the training phase from the evaluation phase.

Table \ref{table:aggregate} shows the average \textbf{MS} (Mutation Score) obtained with the MRs generated by \tool for each subject method.
Such average MS  ranges between 6\% and 86\%.
Furthermore, \tool achieves an average MS higher than 20\% in 18 out of the 23 methods and higher than 50\% in 10 out of the 23 methods. Since the MRs generated by \tool are reusable properties of the methods that can be integrated into any existing test suite, or even implemented as runtime checks, we consider an MS of 20\% to be effective enough, and 50\% to be very effective.

Table~\ref{table:aggregate} shows that the \textbf{PZ} (ratio of MRs without FPs) ranges between 3\% and 90\%. For 15 out of the 23 methods, PZ was no less than 25\%, which indicates that at least one valid MR is generated on average in a single run of \tool with our configuration. Generally, all the methods where \tool achieves a low MS are explained by a similarly low PZ, which might indicate that the dataset of correct test executions used for generating MRs for those methods was not comprehensive enough.

\begin{custombox}{RQ1 -- In summary}
\tool generated valid and effective MRs for the majority of the subject methods.
\end{custombox}

\subsection{RQ2: Test Case Oracle Enhancement}

RQ2 evaluates the degree to which the generated MRs can improve the fault detection capability of test inputs and test case oracles generated by \textsc{Randoop} and \textsc{Evosuite}.
To achieve this, we ran \textsc{Randoop} and \textsc{Evosuite} 12 times with assertions enabled, using the same configuration we used to generate test suites for the MRs with \textsc{Randoop}. 
Columns \textbf{\textsc{Randoop} MS} and \textbf{\textsc{Evosuite} MS} in Table~\ref{table:aggregate} show the mean mutation score (MS) obtained by the \textsc{Randoop} and \textsc{Evosuite} test suites.

Automated test generators like \textsc{Randoop} produce test case oracles that capture the observed behavior of the generated test inputs to expose regression faults as the software evolves.
However, test case oracles only capture the expected behavior for a \emph{specific} input.
Differently, MRs represent universal invariants that can be used to enhance any test suite. 

We can notice that the MS values in Table~\ref{table:aggregate} are often comparable, with a tendency of test case assertions to achieve higher MS values. 
However, the MS values of \textsc{Randoop} and \textsc{EvoSuite}  are obtained by substantially different types of oracles w.r.t. \tool, the former being  test case assertions that predicate on single executions, and the latter  metamorphic relations that hold for all program executions. To assess the degree of complementarity of these different types of oracles, we considered the mutants missed by \textsc{Randoop} or \textsc{Evosuite} and checked whether they are killed by \tool.

Specifically, we automatically enhanced the test suites generated by \textsc{Randoop} and \textsc{Evosuite} by adding  all the MRs generated by \tool. Each MR takes a source input produced by the test generation tools, generates the corresponding follow-up test input, and checks if the output relation is satisfied.
In Table \ref{table:aggregate},
columns \textbf{\tool $\Delta$MS$_R$} and \textbf{\tool $\Delta$MS$_E$} show the percentage of the survived mutants (i.e., those not killed by \textsc{Randoop} or \textsc{Evosuite} test inputs and oracles) that are killed by the enhanced test suite.
We can notice that \tool's MRs are highly complementary to the test case assertions generated by \textsc{Randoop} and \textsc{Evosuite}, as respectively in 7 out of 23 and in 14 out of 23 cases the inclusion of \tool's MRs produce a mutation score improvement ($\Delta$MS $>$ 0).

Let us consider a few examples. 
For \texttt{nextPrime}, \textsc{Randoop} test suites alone kill 80\% of the mutants. 
When this test suite is enhanced with the MRs generated by \tool, the mutation score is 95\%, which means that the enhanced test suites kill 75\% of the remaining mutants, i.e.,  $\Delta$MS$_R$ = 75\% = ( $\frac{95\% - 80\%}{100\% - 80\%}$).
 
The results show a great improvement of the \textsc{Randoop} MS for \texttt{nextPrime}, \texttt{isPrime} and \texttt{gcd} ($\Delta$MS$_R$ between 59\% and 75\%), as well as a significant improvement for \texttt{stirlings} and \texttt{isSorted} ($\Delta$MS$_R$ of 24\% and 25\%) and a small improvement for \texttt{acos} and \texttt{abbreviate} ($\Delta$MS$_R$ of 2\% and 1\%). Although the MRs do not kill any additional mutants for the remaining methods, \textsc{Randoop}'s MS is already 100\% for \texttt{log10}, \texttt{indexOf} and \texttt{truncate}, so improvement is not possible for those methods.

As for \textsc{Evosuite}, the results show a great improvement for the \texttt{nextPrime}, \texttt{difference} and \texttt{repeat} methods ($\Delta$MS$_E$ of 46\%, 50\% and 69\%, respectively), as well as a significant improvement for seven other methods ($\Delta$MS$_E$ between 10\% and 28\%) and a small improvement for four other methods ($\Delta$MS$_E$ between 1\% and 8\%). \textsc{Evosuite}'s MS is 100\% for \texttt{pow}, \texttt{indexOf} and \texttt{meanOf}, making it impossible any improvement in those methods.

\begin{custombox}{RQ2 -- In summary}
\tool generated MRs that enhance the fault detection capability of both \textsc{Randoop} and \textsc{Evosuite}-generated regression test inputs. This shows the complementarity between \tool's MRs and \textsc{Randoop} and \textsc{Evosuite}'s test case assertions.
\end{custombox}

\subsection{RQ3: AutoMR Comparison}

RQ3 compares \tool with the \textsc{AutoMR} \cite{automr} numeric MR generation tool.

\textsc{AutoMR}~\cite{automr} is based on \textsc{MRI}~\cite{2014-zhang-ase}, a search-based approach to automatically generate polynomial MRs. \textsc{MRI} supports MRs in which the input relation is a linear function, and the output relation is either linear or quadratic. 
\textsc{MRI} calculates the optimal coefficients of the linear and quadratic equations using \emph{particle swarm optimization}~\cite{Poli2007}.

\textsc{MRI} uses 1,000 randomly-generated test inputs to filter the MRs that fail for a non-negligible percentage of those inputs.
\textsc{AutoMR}~\cite{automr} addresses several limitations of \textsc{MRI}: \textsc{MRI} does not support MRs with (i) inequality, (ii) more than one input, and (iii) relations of higher degrees than quadratic.

Like \textsc{MRI}, \textsc{AutoMR} uses particle swarm optimization to search for the MR parameters. Similar to \tool, \textsc{AutoMR} generates MRs as pairs of input and output relations, although the latter defines both as arbitrary-degree polynomial relations, and parameterizes both within the same search process. Since \textsc{AutoMR} is an improvement of the approach used by \textsc{MRI}, we compare the effectiveness of \tool against \textsc{AutoMR} only.

We employ the code implemented by the \textsc{AutoMR} authors\footnote{\url{https://github.com/bolzzzz/AutoMR}} in our evaluation. Since this tool is implemented in Python, we implemented a remote function call protocol\footnote{\url{https://github.com/jonayerdi/JavaPythonBridge}} in order to allow this tool to call the experimental subjects, which are Java methods. Since \textsc{AutoMR} calls the subject methods during the search (whereas \tool calls them before) and the remote function call protocol introduces some additional overhead, we did not implement a fixed time budget as with \tool, but instead employed the same configuration used by the \textsc{AutoMR} authors: 3 PSO runs with maximum iterations of 350 for each experiment.

\textsc{AutoMR} has several additional parameters to select for the generated MR, namely: (1) Number of inputs involved, (2) mode of input relation (equality, greater-than, less-than), (3) mode of output relation, (4) degree of input relation (linear, quadratic, etc.), and (5) degree of output relation. We employ the 13 different parameterizations that were defined in the settings from the \textsc{AutoMR} codebase, and also left other configurations unchanged. Same as for \tool, we repeated each run 12 times with different random seeds due to the stochasticity of \textsc{AutoMR}.

As for the experimental subjects, we ran \textsc{AutoMR} on the 10 methods from the Apache Commons Math 3 library. The reason why we evaluated only these methods is that they are purely numeric and contain no variable-length parameters, as it is unclear how \textsc{AutoMR} would be applied to methods that do not fulfill this criteria. In order to convert these subject methods to purely numeric ones, Boolean values (output of the \texttt{isPrime} method) were converted to 0 (false) and 1 (true), and Java Exceptions were converted to 0 outputs.

To evaluate the MRs generated by \textsc{AutoMR}, we performed the same PIT runs that we used for \tool, with the same seed inputs. 
In order to avoid potential implementation issues with the MR evaluation, the \textsc{AutoMR} test suites employed a remote protocol to compute the verdicts in Python, where code derived from  \textsc{AutoMR} was used to provide a pass or fail result. 
Our verdict function employs the same formulas used by \textsc{AutoMR} for evaluating the output relations:

\begin{equation}
\tag{AutoMR equality OR}
    \mathbb{R}_{output} : |\mathbf{Bv}| < 0.05
\end{equation}
\vspace{-16pt}
\begin{equation}
\tag{AutoMR greater-than OR}
    \mathbb{R}_{output} : \mathbf{Bv} > 0
\end{equation}
\vspace{-16pt}
\begin{equation}
\tag{AutoMR less-than OR}
    \mathbb{R}_{output} : \mathbf{Bv} < 0
\end{equation}

Column \textbf{\textsc{AutoMR} MS} of  Table \ref{table:aggregate} shows that \tool achieves a higher mutation score in 9 out of the 10 subject methods for which \textsc{AutoMR} was run. Furthermore, the difference in the method where \tool was outperformed (\texttt{sinh}) is very small, with MSs of 21\% for \tool and 23\% for \textsc{AutoMR}. Note that \textsc{AutoMR} produced no output for 5 out of the 10 subject methods, as all the generated MRs were filtered out during the final redundant MR removal phase, hence the $0.00$ values in the \textsc{AutoMR} MS column.

Column \textbf{\tool $\Delta$MS$_A$} shows the percentage of the survived mutants (i.e., those not killed by \textsc{AutoMR} MRs) that are killed by \tool MRs. These results show that \tool MRs could identify several additional mutants that could not be detected with \textsc{AutoMR} in all the subject methods, with $\Delta$MS$_A$ values ranging between 17\% and 72\% for the subject methods where both tools produced effective MRs.

\begin{custombox}{RQ3 -- In summary}
\tool generated MRs that achieved a higher mutation score than those generated by \textsc{AutoMR} for 9 out of 10 subject methods. The MRs generated by \tool kill more mutants than the MRs generated by \textsc{AutoMR} in all subject methods.
\end{custombox}

\subsection{RQ4: Filtering}
RQ4 evaluates the effectiveness of the filtering process in detecting invalid MRs.

Table~\ref{table:aggregate} shows that the \textbf{PZ} values are not greater than 86\% for any of the methods, so at least 14\% of the generated MRs are always filtered. In fact, more than 50\% of the MRs are filtered in 13 out of 23 methods. On the one hand, we observe that the PZ is particularly low in methods involving boolean inputs or outputs. This might indicate that methods with boolean inputs or outputs can hold properties that seem valid for a large variety of inputs, but do not actually generalize for all possible inputs. Furthermore, since booleans only have two possible values, \tool may generate more specific output relations, which are more prone to FPs. On the other hand, we also noticed that the filtering process eliminated most MRs for the \texttt{acos} and \texttt{log10} methods, which involve floating-point arithmetic. A possible reason could be that floating-point arithmetic has special values such as NaN or Infinity, which may invalidate some MRs which would be valid for the cases where all inputs and outputs are finite values.

To provide more insights on the contribution of OASIs for identifying invalid MRs, Table~\ref{table:aggregate} also shows the \textbf{PZO} metric, the ratio of MRs for which OASIs does not find FPs. Note that only the MRs which yield no FPs with all the Randoop-generated test inputs are passed to OASIs, hence PZO is a ratio calculated over the MRs which already passed the Randoop filter. The results shows that OASIs can identify additional FPs not found by Randoop in 17 of the 23 subject methods. For \texttt{log10}, \texttt{sinh} and \texttt{indexOf}, OASIs identified FPs in more than 40\% of the MRs that passed the Randoop filter, indicating that Randoop regression test suites alone may be insufficient to discard invalid MRs in some cases.

\begin{custombox}{RQ4 -- In summary}
\tool's filtering process is effective in filtering MRs that do not generalize well on unseen inputs, and OASIs can successfully identify invalid MRs that are not identified with Randoop alone.
\end{custombox}

\subsection{Threats to Validity}

\noindent \textbf{External validity.} An external threat in our evaluation relates to the generalization of the results. We mitigated this threat by selecting 23 diverse functions from three different codebases. 
Moreover, we evaluated \tool with unseen test inputs and seeded faults generated with different tools from the ones used to generate the MRs.

\noindent \textbf{Internal validity.} 
Internal validity threats arise from errors in the measurements or the implementation of \tool. We mitigate this threat by manually inspecting the output and internal behavior of \tool for a few sample runs.

\noindent \textbf{Conclusion validity.} The approach and evaluation methods are inherently stochastic. To account for this, we ran each technique 12 times and evaluated each generated MR with 12 different test suites, resulting in $12 \times 12 = 144$ samples.

\section{Related Work}
\smallskip
\textbf{Automatic test generation.}
Test generators are often meant for regression testing scenarios, and therefore capture the implemented behavior of the software under test in order to expose new faults as the software evolves~\cite{comparison,jin2010automated,fraser2011evosuite,pacheco2007randoop}.
Similarly, \tool also captures the implemented behavior of the system. In fact, \tool's FN-based guidance is effectively very similar to the mutation-guided assertion generation used by Evosuite~\cite{fraser2011evosuite}, since the states used to compute FNs in \tool are obtained with mutation testing. Unlike test generators, however, \tool produces universal properties of the system (MRs) that can be tested for any input, instead of input-specific assertions.

\smallskip
\noindent \textbf{Metamorphic Testing.} 
Most metamorphic testing approaches assume the availability of MRs~\cite{2016-segura-tse}. For instance, much work has been done to predict whether an MR picked from a predefined list is suitable 
for a given program (e.g., ~\cite{Duque2022,hardin2018using,rahman2018predicting,zhang2017rbf,rahman2020mrpredt,dwarakanath2018identifying,Nair2019}). Differently, \tool aims to automatically generate new MRs for a given program, which remains a less studied problem~\cite{2016-segura-tse,2013-hui-wcse,2017-chen-cs}. We now discuss the most related work to the generation of new MRs.

\smallskip
\noindent \emph{ML-based approaches.} Researchers investigated the use of machine-learning to predict whether specific types of MRs~\cite{murphy2008properties} hold for a given method~\cite{2013-kanewala-issre,2015-kanewala-stvr,zhang2017rbf}. 
These approaches only predict whether pre-defined types of MRs hold for a method. Developers are expected to derive an executable MR using this information~\cite{2013-kanewala-issre}.
Conversely, \tool automatically generates executable MRs.

\smallskip
\noindent \emph{NLP-based approaches.}
\textsc{MeMo}~\cite{BlasiGEPC21} automatically derives  equivalence MRs from the JavaDoc.
The quality of derived MRs, by design, strictly depend on the completeness and correctness of the available documentation. Differently, \tool does not rely on documentation and it is not limited to MRs expressed as equivalences.

\smallskip
\noindent \emph{Search-based approaches.}
\textsc{MRI}~\cite{2014-zhang-ase} and \textsc{AutoMR} \cite{automr} are search-based approaches which employ particle-swarm optimization~\cite{Poli2007} to parameterize polynomial input and output relations.

Same as \tool, \textsc{MRI} and \textsc{AutoMR} use  search-based algorithms to generate MRs for numerical programs. 
However, the fitness functions of \textsc{MRI} and \textsc{AutoMR} take into account only the number of FPs. 
\tool's fitness functions also consider the number of FNs, which is crucial to obtain MRs that are effective at exposing software faults.  
Moreover, \tool is not limited to polynomial MRs, but can also operate on logical predicates and ordered sequences.
We have conducted a comparison between \textsc{AutoMR} and \tool with numerical subject methods only, since non-numeric functions are not supported by \textsc{AutoMR}.

\gassert~\cite{gassert} and \textsc{EvoSpex}~\cite{evospex} generate oracles with evolutionary algorithms driven by both FPs and FNs. However, they target program assertions and invariants, respectively, and they cannot be easily adapted to target MRs.
Moreover, their representation of FPs and FNs is ill-suited for MRs as discussed in Section~2.
Ayerdi et al. proposed \textsc{GAssertMRs}~\cite{gassertmrs}, which adapts \textsc{GAssert}~\cite{gassert,gassert-tool} for generating MRs for CPS, considering, in particular, an industrial system of elevators as case study. 
However, \textsc{GAssertMRs} supports domain-specific, system-level MRs that relate configuration variables for system level test scenarios (e.g., the number of elevators or passengers in each floor) with system level quality metrics (e.g., average waiting time). On the contrary, \tool operates at the unit level and generates MRs that predicate on local variables of the method under test.

\section{Conclusions and Future Work}
This paper presented \tool, the first generator of  metamorphic relations capable of minimizing at the same time the false positive rate (associated with false alarms) and the false negative rate (associated with missed faults). It includes a filtering phase that further eliminates   MR oracles prone to residual false positives. 

Our empirical results show that our evolutionary approach generates useful and non-trivial MRs, which can expose faults simulated by mutants in the majority of the analyzed subjects. The metamorphic oracle produced by \tool has  been shown to be highly complementary to the test case assertions automatically generated by test generators like \textsc{Randoop} and \textsc{EvoSuite}. 
In fact, when the MRs generated by \tool are added to \textsc{Randoop} and \textsc{EvoSuite}'s test cases, they increase the fault detection capability of the latter tools by a substantial amount. We have also evaluated the usefulness of the filtering phase, showing in particular that the oracle validator OASIs gives a fundamental contribution to the elimination of candidate oracles that trigger false alarms.

Future works aim at increasing the effectiveness and applicability of \tool. We discuss the most promising ones.

\textbf{Extending \tool to handle complex types}. 
While the general design of \tool allows for arbitrarily complex expressions and type systems, our current implementation for Java methods only supports Boolean, numeric and ordered-sequence types for variables and operations.
More research is needed to support complex types (e.g. user-defined classes) and their operations in the generated MRs. 

\textbf{Improving MRs readability}. We found that several MRs generated by \tool could be simplified to increase their readability.
For example, by removing tautologies, reordering the variables, or replacing sub-expressions with equivalent but shorter ones. An important future work is to investigate post-processing analyses to simplify the generated MRs.

%
%
%
%




\ifCLASSOPTIONcompsoc
  \section*{Acknowledgments}
\else
  \section*{Acknowledgment}
\fi

This work was partially founded by the Basque Government through their Elkartek program (EGIA project, ref. KK-2022/00119). Jon Ayerdi and Aitor Arrieta are part of the Software and Systems Engineering research group of Mondragon Unibertsitatea (IT1519-22), supported by the Department of Education, Universities and Research of the Basque Country. This work was partially supported by the H2020 project PRECRIME, funded under the ERC Advanced Grant 2017 Program (ERC Grant Agreement n. 787703). Gunel Jahangirova has been partially supported by the UKRI
Trustworthy Autonomous Systems Node in Verifiability, Grant Award Reference EP/V026801/2.

\ifCLASSOPTIONcaptionsoff
  \newpage
\fi



%
\bibliographystyle{IEEEtran}
\bibliography{main}

%





\end{document}